# Arrival time in quantum field theory

Zhi-Yong Wang[1], Cai-Dong Xiong[1], Bing He[2]

[1]*School of Optoelectronic Information, University of Electronic Science and Technology of China, Chengdu 610054, CHINA*
[2]*Department of Physics and Astronomy, Hunter College of the City University of New York, 695 Park Avenue, New York, NY 10065, USA*

**Abstract**

Via the proper-time eigenstates (event states) instead of the proper-mass eigenstates (particle states), free-motion time-of-arrival theory for massive spin-1/2 particles is developed at the level of quantum field theory. The approach is based on a position-momentum dual formalism. Within the framework of field quantization, the total time-of-arrival is the sum of the single *event-of-arrival* contributions, and contains zero-point quantum fluctuations because the clocks under consideration follow the laws of quantum mechanics.

## 1. Introduction

Free-motion time-of-arrival theories have been developed at the level of nonrelativistic quantum mechanics [1-5]. In Ref. [6] we have developed relativistic free-motion time-of-arrival theory for massive spin-1/2 particles. Relativistic quantum mechanics is only a transitional theory to quantum field theory, then for completeness, it is necessary to study arrival time at the level of quantum field theory. In traditional quantum field theory, energy and momentum are dynamical variables while time and space coordinates are parameters. For our goal, our study is based on the event states satisfying the 4D spacetime interval relation $T^2 = x^2 + \tau^2$ instead of the particles states satisfying the mass-shell relation $E^2 = m^2 + p^2$, where the symmetry of a physical system under a shift of zero-energy reference point is applied. In this article the natural units of measurement ($\hbar = c = 1$) is applied, the space-time metric tensor is $g^{\mu\nu} = \text{diag}(1,-1,-1,-1)$, $\mu,\nu = 0,1,2,3$.

The first attempt to develop the field-quantized theory of time-of-arrival has been presented by A. D. Baute *et al* [7], in which the authors proposed a prescription for computing the density of arrivals of particles for multiparticle states both in free and interacting case, by applying the concept of the crossing state and the formalism of field quantization. Our work is different from theirs in the following aspects: (1) Their work is based on the nonrelativistic quantum theory, while our work is based on the relativistic quantum theory with a correct nonrelativistic limit; (2) Their work is only valid for the scalar particles, while our work is valid for both the spin-1/2 particles and the scalar particles by ignoring the spin degrees of freedom; (3) Their work is based on multiparticle states without a systematic quantum-field-theory framework, while our work is based on *multievent* states rather than multiparticle states, and a systematic quantum-field-theory framework is constructed via a position-momentum dual formalism.

**2. Free-motion time-of-arrival operator of spin-1/2 particles and its eigenfunctions**

To present a self-contained argument, let us first mention some of the contents presented in Ref. [6]. Let $\gamma^\mu$'s ($\mu = 0,1,2,3$) denote the 4×4 Dirac matrices satisfying the algebra $\gamma^\mu \gamma^\nu + \gamma^\nu \gamma^\mu = 2g^{\mu\nu}$, $\boldsymbol{\alpha} = (\alpha_1, \alpha_2, \alpha_3)$ denote a matrix vector with the components $\alpha_i = \beta \gamma^i$ ($i = 1,2,3$), where $\beta = \gamma^0$. A free spin-1/2 particle of rest mass $m$ has the Hamiltonian $\hat{H} = \boldsymbol{\alpha} \cdot \hat{\boldsymbol{p}} + \beta m$. For simplicity, we choose a coordinate system with its x-axis being parallel to the momentum of the particle, such that the four-dimensional (4D) momentum of the particle becomes $p^\mu = (E, p, 0, 0)$, where $E^2 = p^2 + m^2$. For our purpose, we assume that whenever $p \neq 0$, i.e., $E^2 \neq m^2$, this condition presents no problem for our issues. In the present case, the Hamiltonian becomes $\hat{H} = \alpha_1 \hat{p} + \beta m$, where $\hat{p} = -i\partial/\partial x$, and the Dirac equation can be converted into

$$i\partial \psi(t,x)/\partial t = (\alpha_1 \hat{p} + \beta m)\psi(t,x). \tag{1}$$

Here the 2D form is embedded in the 4D space-time background, and then the spin degrees of freedom should still be taken into account. The classical expression for the relativistic arrival time at the origin $x_0 = 0$ of the freely moving spin-1/2 particle having position $x$ and uniform velocity $p/E$, is $T = -x(E/p)$ ($\hbar = c = 1$, note that $x = \Delta x = x - x_0$ is a space interval). There are many quantization schemes. For convenience, we choose Weyl's prescription as our quantization scheme, then the transition from the classical expression $T = -x(E/p)$ to a quantum-mechanical one requires us to symmetrize the product between $x$, $E$ and $p$, and replace $x$, $E$ and $p$ with their quantum-mechanical operators, respectively, in such a way one can obtain the quantum-mechanical form of $T = -x(E/p)$ [6, 8]:

$$\hat{T}_{\text{Dirac}}(\hat{x}, \hat{p}) = -(\alpha_1 \hat{x} + \beta \hat{\tau}), \tag{2}$$

where $-\hat{\tau} = -m(\hat{p}^{-1}\hat{x} + \hat{x}\hat{p}^{-1})/2$ is the nonrelativistic time-of-arrival operator that has been studied thoroughly in previous literatures [1-5], and it plays the role of proper time-of-arrival operator [6]. The operator $\hat{T}_{\text{Dirac}}$ given by Eq. (2) represents the relativistic time-of-arrival operator of the Dirac particles, it canonically conjugates to the Hamiltonian operator $\hat{H} = \alpha_1 \hat{p} + \beta m$, i.e., $[\hat{H}, \hat{T}_{\text{Dirac}}] = i$. Owing to the particle-antiparticle symmetry, $\hat{T}_{\text{Dirac}}$ has a self-adjoint extension without contradicting Pauli's argument [6]. In the momentum representation, Eq. (2) becomes

$$\hat{T}_{\text{Dirac}}(\hat{x}, \hat{p}) = (1/p)(\alpha_1 p + \beta m)(-i\partial/\partial p) + i\beta m/2p^2. \tag{3}$$

Assume that the eigenequation of $\hat{T}_{\text{Dirac}}(\hat{x},\hat{p})$ is

$$\hat{T}_{\text{Dirac}}(\hat{x},\hat{p})\phi(p) = T\phi(p), \tag{4}$$

one can prove that the eigenvalues and eigenfunctions of $\hat{T}_{\text{Dirac}}(\hat{x},\hat{p})$ are, respectively (see **Appendix A**),

$$\phi_{xs}(p) = (p^2/p^2+m^2)^{1/4} u(p,s)\exp(-ipx)/(2\pi)^{1/2}, \text{ for } T = -T_x, \tag{5}$$

$$\phi_{-xs}(p) = (p^2/p^2+m^2)^{1/4} v(p,s)\exp(ipx)/(2\pi)^{1/2}, \text{ for } T = T_x, \tag{6}$$

where $s = \pm 1/2$, $T_x = xE_p/p$, $E_p = \sqrt{p^2+m^2}$, and

$$u(p,s) = \sqrt{\frac{m+E_p}{2E_p}}\begin{pmatrix}\eta_s \\ \frac{\sigma_1 p}{m+E_p}\eta_s\end{pmatrix}, \quad v(p,s) = \sqrt{\frac{m+E_p}{2E_p}}\begin{pmatrix}\frac{\sigma_1 p}{m+E_p}\eta_s \\ \eta_s\end{pmatrix}. \tag{7}$$

where

$$\sigma_1 = \begin{pmatrix}1 & 0 \\ 0 & -1\end{pmatrix}, \quad \eta_{1/2} = \begin{pmatrix}1 \\ 0\end{pmatrix}, \quad \eta_{-1/2} = \begin{pmatrix}0 \\ 1\end{pmatrix}. \tag{8}$$

For example, to examine whether Eq. (5) satisfies Eq. (4), one can apply the formulae:

$$\frac{\partial}{\partial p}(\frac{p^2}{p^2+m^2})^{1/4} = \frac{m^2}{2E_p^2}\frac{1}{p}(\frac{p^2}{p^2+m^2})^{1/4}, \quad \frac{\partial}{\partial p}u(p,s) = (\frac{m^2}{2E_p^2})\alpha_1\beta u(p,s). \tag{9}$$

For our purpose, substituting $E_p = T_x p/x$ and $\tau = xm/p$ into Eq. (7), one has

$$u(p,s) = \sqrt{\frac{\tau+T_x}{2T_x}}\begin{pmatrix}\eta_s \\ \frac{\sigma_1 x}{\tau+T_x}\eta_s\end{pmatrix} \equiv \zeta(x,s), \quad v(p,s) = \sqrt{\frac{\tau+T_x}{2T_x}}\begin{pmatrix}\frac{\sigma_1 x}{\tau+T_x}\eta_s \\ \eta_s\end{pmatrix} \equiv \xi(x,s). \tag{10}$$

Using $\tau = xm/p$ one has $p^2/p^2+m^2 = x^2/x^2+\tau^2$, then in terms of $\zeta(x,s)$ and $\xi(x,s)$ defined by Eq. (10), Eqs. (5) and (6) can be rewritten as, respectively,

$$\phi_{xs}(p) = (x^2/x^2+\tau^2)^{1/4}\zeta(x,s)\exp(-ipx)/(2\pi)^{1/2}, \text{ for } T = -T_x, \tag{11}$$

$$\phi_{-xs}(p) = (x^2/x^2+\tau^2)^{1/4}\xi(x,s)\exp(ipx)/(2\pi)^{1/2}, \text{ for } T = T_x, \tag{12}$$

Here one can obtain Eqs. (11) and (12) just by substituting $E_p = T_x p/x$ and $\tau = xm/p$

into Eqs. (5) and (6), and then Eqs. (11) and (12) are completely equivalent to Eqs. (5) and (6), respectively. Just as a Hamiltonian operator such as $\hat{H}(\hat{x},\hat{p})$, the arrival time operator $\hat{T}_{\text{Dirac}}(\hat{x},\hat{p})$ is a twovariable function of position and momentum. Likewise, $T = T(x,p)$ is a twovariable function of $x$ and $p$, where $x$ and $p$ are two independent variables such that $\partial x/\partial p = 0$ and $\partial p/\partial x = 0$, that is, $x$ is not explicitly dependent on $p$ and vice versa. Consider that $T^2 = T_x^2 = x^2 + \tau^2 = (xE_p/p)^2$, in the following we take $T_x = \sqrt{x^2 + \tau^2}$, which has no effect on our final results.

**3. Energy shift equation and proper-time eigenstates**

Let us introduce an *energy parameter* $\varepsilon$ with the dimension of energy and independent of the momentum $p$ (i.e., $\partial \varepsilon/\partial p = 0$). Contrary to the Hamiltonian $\hat{H} = \alpha_1 \hat{p} + \beta m$ with the spectrum $\mathcal{R}_m = (-\infty,-m)\bigcup(m,+\infty)$, here one has $\varepsilon \in (-\infty,+\infty)$. For example, one can write $\varepsilon = E + \varepsilon_0$, where $\varepsilon_0$ is called the zero-energy reference point and behaves as a constant and uniform potential. Let $\phi(\varepsilon,p) = \exp(\mathrm{i}T\varepsilon)\phi(p)$, one can rewrite Eq. (4) as

$$-\mathrm{i}\partial\phi(\varepsilon,p)/\partial\varepsilon = \hat{T}_{\text{Dirac}}\phi(\varepsilon,p). \qquad (13)$$

Eq. (13) is equivalent to Eq. (4), but it is expressed as a dual form of the Schrödinger equation $\mathrm{i}\partial\psi(t,x)/\partial t = \hat{H}\psi(t,x)$. As we know, the Schrödinger equation describes the time evolution of the state $\psi(t,x)$, and all the related conserved quantities are invariant in the evolution process. Likewise, we call Eq. (13) *energy shift equation*. In our case, physical observables involve energy differences and not the absolute value of energies, and do not depend on the choice of zero-energy reference points, then one can introduce the energy parameter $\varepsilon$ to describe the degree of freedom related to the shift of a zero-energy reference point, and apply Eq. (13) to describe the energy-shift evolution of the state $\phi(\varepsilon, p)$, for the moment all physical observables are invariant in this evolution process. Moreover, according to Ref. [9], one can call $\hat{T}_{\text{Dirac}}$ "time-Hamiltonian", or, seeing that a Hamiltonian can be called energy function, one can also call $T = -x(E/p)$ "time function" while call $\hat{T}_{\text{Dirac}}$ "time function operator" [10].

Using $\phi(\varepsilon, p) = \exp(iT\varepsilon)\phi(p)$, Eqs. (4), (11) and (12), one can easily show that the elementary solutions of Eq. (13) are

$$\phi_{xs}(\varepsilon, p) = (x^2/x^2 + \tau^2)^{1/4} \zeta(x,s) \exp[-i(\varepsilon T_x + px)]/(2\pi)^{1/2}, \text{ for } T = -T_x, \quad (14)$$

$$\phi_{-xs}(\varepsilon, p) = (x^2/x^2 + \tau^2)^{1/4} \xi(x,s) \exp[i(\varepsilon T_x + px)]/(2\pi)^{1/2}, \text{ for } T = T_x. \quad (15)$$

However, the states given by Eqs. (14) and (15) are not the proper-time eigenstates, in which the representation of the time-of-arrival operator is $\hat{T}_{\text{Dirac}} = -(\alpha_1 \hat{x} + \beta \hat{\tau})$ rather than $\hat{T}_{\text{Dirac}} = -(\alpha_1 \hat{x} + \beta \tau)$, namely, the proper-time term is an operator rather than a c-number. For convenience, we will apply the representation of $\hat{T}_{\text{Dirac}}$ in the proper-time eigenstates, i.e., $\hat{T}_{\text{Dirac}} = -(\alpha_1 \hat{x} + \beta \tau)$. Owing to the fact that the proper-time eigenstates are no longer the proper-mass eigenstates, once one takes $\hat{T}_{\text{Dirac}} = -(\alpha_1 \hat{x} + \beta \tau)$, the Hamiltonian operator is no longer $\hat{H} = \alpha_1 \hat{p} + \beta m$, but rather $\hat{H} = \alpha_1 \hat{p} + \beta \hat{m}$, where $\hat{m}$ is the proper-mass operator. For example, from $\tau = xm/p$ one can obtain $\hat{m} = (\tau/2)[\hat{p}(1/\hat{x}) + (1/\hat{x})\hat{p}]$ via Weyl's quantization scheme, and one can easily prove that the commutation relation $[\hat{T}_{\text{Dirac}}, \hat{H}] = -i$ is still valid for $\hat{T}_{\text{Dirac}} = -(\alpha_1 \hat{x} + \beta \tau)$ and $\hat{H} = \alpha_1 \hat{p} + \beta \hat{m}$, then $\hat{T}_{\text{Dirac}}$ and $\hat{H}$ keep forming a canonical conjugate pair.. For the moment, the proper time $\tau$ is a c-number variable not explicitly dependent on $p$ (i.e., $\partial \tau/\partial p = 0$), and Eq. (13) becomes

$$-i \partial \varphi(\varepsilon, p)/\partial \varepsilon = -(\alpha_1 \hat{x} + \beta \tau)\varphi(\varepsilon, p). \quad (16)$$

One can prove that elementary solutions of Eq. (16) are

$$\varphi_{xs}(\varepsilon, p) = \zeta(x,s)\exp[-i(\varepsilon T_x + px)]/(2\pi)^{1/2}, \text{ for } T = -T_x, \quad (17)$$

$$\varphi_{-xs}(\varepsilon, p) = \xi(x,s)\exp[i(\varepsilon T_x + px)]/(2\pi)^{1/2}, \text{ for } T = T_x. \quad (18)$$

There are quantum interferences between different quantum events, then it is meaningful to discuss the general solution of Eq. (16), which can be written as a linear combination of the elementary solutions given by Eqs. (17) and (18), that is:

$$\varphi(\varepsilon, p) = \frac{1}{\sqrt{M}} \sum_{x,s} [a(x,s)\varphi_{xs}(\varepsilon, p) + b^\dagger(x,s)\varphi_{-xs}(\varepsilon, p)]. \quad (19)$$

where $M$ satisfies $\lim_{M\to +\infty}(1/M)\int_M dp = 1$, $a(x,s)$ and $b^\dagger(x,s)$ (the hermitian conjugate of $b(x,s)$) are expansion coefficients. Here the integral $\int(\cdot)dx$ is expressed as the form of the discrete sum $\sum_x (\cdot)$.

**4. Position-momentum dual formalism and the field quantization of arrival time**

In Eq. (1), the state $\psi(t,x) = \psi(t)$ describes "particle state" (proper-mass eigenstates) satisfying the mass-shell relation $E^2 = p^2 + m^2$. Analogically, we refer to the state $\varphi(\varepsilon, p) = \varphi(\varepsilon)$ in Eq. (16) as "event states" (proper-time eigenstates) which satisfies the spacetime interval relation $T^2 = x^2 + \tau^2$ (note that $x = \Delta x = x - x_0$ is a space interval). The time-of-arrival operator $\hat{T}_{\text{Dirac}} = -\alpha_1 \hat{x} - \beta \tau$ is to $T^2 = x^2 + \tau^2$ as the Hamiltonian operator $\hat{H} = \alpha_1 \hat{p} + \beta m$ is to $E^2 = p^2 + m^2$, which shows us a duality between the position and momentum space. To study quantum-field-theory arrival time we will resort to the position-momentum dual formalism, which is based on "event states" (proper-time eigenstates) instead of particle states (proper-mass eigenstates). Some previous attempts of describing event eigenstates can be found in Ref. [11-13], but they are based on the usual quantum theory, and different from our duality formalism with the following dual relations:

$$\begin{cases} T^2 = x^2 + \tau^2 \leftrightarrow E^2 = p^2 + m^2 \\ \hat{T} = -\alpha_1 \hat{x} - \beta\tau \leftrightarrow \hat{H} = \alpha_1 \hat{p} + \beta m \\ -i\partial\varphi(\varepsilon)/\partial\varepsilon = \hat{T}\varphi(\varepsilon) \leftrightarrow i\partial\psi(t)/\partial t = \hat{H}\psi(t) \end{cases}. \quad (20)$$

In general, if a physical quantity $Q$ satisfies $\partial Q/\partial\lambda = 0$ for a parameter $\lambda$, we call $Q$ a *generalized conserved quantity* with respect to the parameter $\lambda$ (i.e., $Q$ has a value constant in $\lambda$). In particular, if $\lambda$ is the energy parameter $\varepsilon$ introduced before, it implies that the generalized conserved quantity $Q$ is invariant under an energy shift, and then do not depend on the choice of zero-energy reference point. For example, it has no effect on a freely moving spin-1/2 particle to choose a new zero-energy reference point, because it is equivalent to putting the particle into a constant and uniform potential field. As a result, the time-of-arrival $T = -x(E/p)$ (or the operator $\hat{T} = \hat{T}_{\text{Dirac}}$) is such a generalized conserved quantity. In other words, under an energy-shift transformation, the invariance of a system implies the independence of zero-energy reference point, i.e., the energy-shift symmetry of the system.

Related to what is mentioned above, a dual counterpart of the traditional mechanics can be introduced, which is different from the momentum-space representation of the traditional mechanics. Let $A(k,q,t)$ be an action, where $q = (q_1, q_2..., q_n)$ and $k = (k_1, k_2..., k_n)$ are the generalized coordinates and momenta, respectively. By the Jacobi-Hamilton equation

$$\partial A/\partial t + H = 0, \quad (21)$$

one can define the Hamiltonian $H$. The Hamilton equations are

$$\begin{cases} \dfrac{\partial k_i}{\partial t} = -\dfrac{\partial H}{\partial q_i} = \dfrac{\partial^2 A}{\partial q_i \partial t} = \dfrac{\partial}{\partial t}(\dfrac{\partial A}{\partial q_i}) \\ \dfrac{\partial q_i}{\partial t} = \dfrac{\partial H}{\partial k_i} = -\dfrac{\partial^2 A}{\partial k_i \partial t} = -\dfrac{\partial}{\partial t}(\dfrac{\partial A}{\partial k_i}) \end{cases}, \quad i = 1, 2, \ldots n. \tag{22}$$

Then

$$k_i = \partial A/\partial q_i + C_{1i}, \quad q_i = -\partial A/\partial k_i + C_{2i}, \quad i = 1, 2, \ldots n, \tag{23}$$

where $C_{1i}$ and $C_{2i}$ ($i = 1, 2, \ldots n$) are constants. For simplicity, let $C_{1i} = C_{2i} = 0$ ($i = 1, 2, \ldots n$), then

$$q_i = -\partial A/\partial k_i, \quad k_i = \partial A/\partial q_i, \quad i = 1, 2, \ldots n. \tag{24}$$

In terms of the energy parameter $\varepsilon$, let us introduce a time function $T$ as follows:

$$T \equiv -\partial A/\partial \varepsilon. \tag{25}$$

Eq. (25) is the dual counterpart of Eq. (21). Using Eqs. (24) and (25), one has

$$\begin{cases} \dfrac{\partial T}{\partial q_i} = -\dfrac{\partial}{\partial \varepsilon}(\dfrac{\partial A}{\partial q_i}) = -\dfrac{\partial k_i}{\partial \varepsilon} \\ \dfrac{\partial T}{\partial k_i} = -\dfrac{\partial}{\partial \varepsilon}(\dfrac{\partial A}{\partial k_i}) = \dfrac{\partial q_i}{\partial \varepsilon} \end{cases}, \quad i = 1, 2, \ldots n. \tag{26}$$

That is

$$\begin{cases} \partial k_i/\partial \varepsilon = -\partial T/\partial q_i \\ \partial q_i/\partial \varepsilon = \partial T/\partial k_i \end{cases}. \tag{27}$$

Eq. (27) is the dual counterpart of the Hamilton equations (22).

As we know, the Hamiltonian operator $\hat{H}$ is the generator of time translation transformations. Likewise, we will show $\hat{T} = \hat{T}_{\text{Dirac}}$ is the generator of energy-shift transformations. To do so, let us define $\bar{\varphi}(\varepsilon, p) \equiv \varphi^{\dagger}(\varepsilon, p)\gamma^0$. In terms of $\varphi(\varepsilon, p)$ and $\bar{\varphi}(\varepsilon, p)$ we define an action ($A$, say) as follows:

$$A = \int \bar{\varphi}(\varepsilon, p)[\gamma^0 \text{i}(\partial/\partial \varepsilon) - \gamma^1 \hat{x} - \tau]\varphi(\varepsilon, p) \text{d}p \text{d}\varepsilon. \tag{28}$$

Obviously, the quantity $A$ has the dimension of action. One can prove that Eq. (16) can be derived via the following variational principle: let the system occupy, at $\varepsilon = \varepsilon_1$ and $\varepsilon = \varepsilon_2$, momentums denoted by $p = p_1$ and $p = p_2$, then the system changes between these momentums in such a way that the action integral given by Eq. (28) takes the least possible value. In the present case, using Eq. (28) we define a generalized Lagrange density as follows:

$$\Gamma = \bar{\varphi}(\varepsilon, p)[\gamma^0 i(\partial/\partial\varepsilon) - \gamma^1 \hat{x} - \tau]\varphi(\varepsilon, p). \tag{29}$$

In the two-dimensional form, the generalized Lagrange density $\Gamma$ has the dimension of $[\text{length}]^2$ rather than that of $[\text{length}]^{-2}$, which is a dual counterpart of the usual Lagrange density. As we know, Eq. (1) is equivalent to the corresponding Hamilton equations. Likewise, let $\dot{\varphi} \equiv \partial\varphi/\partial\varepsilon$, taking $\varphi(\varepsilon, p)$ as the generalized coordinate with its canonically conjugate momentum being $\pi(\varepsilon, p) = \partial\Gamma/\partial\dot{\varphi} = i\bar{\varphi}(\varepsilon, p)\gamma^0$, we define a time function density as (being a dual counterpart of Hamiltonian density)

$$T^\circ = \pi\dot{\varphi} - \Gamma = \varphi^\dagger(\varepsilon, p)(\alpha_1 \hat{x} + \beta\tau)\varphi(\varepsilon, p). \tag{30}$$

For the moment, Eq. (27) becomes

$$\begin{cases} \partial\pi/\partial\varepsilon = -\partial T^\circ/\partial\varphi \\ \partial\varphi/\partial\varepsilon = \partial T^\circ/\partial\pi \end{cases}. \tag{31}$$

It is easy to obtain Eq. (16) from the second equation of Eq. (31), i.e., Eq. (16) is equivalent to $\partial\varphi/\partial\varepsilon = \partial T^\circ/\partial\pi$. Further, one can prove that $\Gamma$ given by Eq. (29) is invariant under the energy-shift transformation $\varepsilon \to \varepsilon' = \varepsilon + a$, where the corresponding generator is $\hat{T}_{\text{Dirac}} = -(\alpha_1 \hat{x} + \beta\tau)$, and the generalized conserved charge (i.e., the one being independent of the energy parameter $\varepsilon$) is

$$T_{\text{Dirac}} = \int \varphi^\dagger(\varepsilon, p)\hat{T}_{\text{Dirac}}\varphi(\varepsilon, p)dp = \int T^\circ dp. \tag{32}$$

Now, let $a^\dagger(x, s)$ and $a(x, s)$ represent the creation and annihilation operators of the electron's arrival events, while $b^\dagger(x,s)$ and $b(x,s)$ represent the ones of the positron's arrival events. Correspondingly, let $N(x,s) = a^\dagger(x,s)a(x,s)$ and $N'(x,s) = b^\dagger(x,s)b(x,s)$ are the event number operators related to the arrival events of electrons and positrons, respectively. Assume that these creation and annihilation operators satisfy the following anti-commutation relations:

$$\{a^\dagger(x,s), a(x',s')\} = \{b^\dagger(x,s), b(x',s')\} = \delta_{xx'}\delta_{ss'}, \qquad (33)$$

and all the others vanish. Eq. (33) implies that the generalized coordinate $\varphi(\varepsilon, p)$ and its conjugate momentum $\pi(\varepsilon, p)$ satisfy the following anti-commutation relations

$$\begin{cases} \{\varphi(\varepsilon, p), \pi(\varepsilon, p')\} = i\delta(p-p') \\ \{\varphi(\varepsilon, p), \varphi(\varepsilon, p')\} = \{\pi(\varepsilon, p), \pi(\varepsilon, p')\} = 0 \end{cases}. \qquad (34)$$

Using Eqs. (10), (17)-(19), (32) and (33), one has

$$T_{\text{Dirac}} = \sum_{x,s}[a^\dagger(x,s)a(x,s) + b^\dagger(x,s)b(x,s) - 1]T_x. \qquad (35)$$

Eq. (35) is the field-quantized expression of arrival time, it implies that $T_{\text{Dirac}}$ represents the total time-of-arrival, i.e., the sum of all the single event-of-arrival contributions, where a single event-of-arrival contributes to a time-of-arrival $T_x = \sqrt{x^2 + \tau^2}$. Then, in our case, the additive character of time of arrivals is realized via event states rather than particle states. Note that $T_{\text{Dirac}}$ as a Fock-space operator, its operator property is entirely carried by the creation and annihilation operators of arrival events. Moreover, Eq. (35) implies that, even if there is not any event-of-arrival, the arrival time contains zero-point quantum fluctuations. In fact, when the time defined by a classical clock takes the value $t$, read the classical clock, of course the result is $t$ without any fluctuation. However, in our case, the time takes the value $T$, read a quantum clock that follows the laws of quantum mechanics, and then the result has quantum fluctuations. Here the quantum clock that indicates its time as a function of the distance is defined by the spin-1/2 particle and a "screen" (position-of-arrival).

## 5. Conclusions

Relativistic free-motion time-of-arrival operator for massive spin-1/2 particles, i.e., $\hat{T}_{\text{Dirac}}(\hat{x}, \hat{p}) = -(\alpha_1 \hat{x} + \beta \hat{\tau})$, can be obtained via Weyl's quantization scheme [6]. To develop the free-motion time-of-arrival theory for massive spin-1/2 particles to the level of quantum field theory, we resort to a position-momentum dual formalism by means of event states (satisfying the 4D spacetime interval relation $T^2 = x^2 + \tau^2$), instead of particle states (satisfying the mass-shell relation $E^2 = m^2 + p^2$), which is based on the symmetry of our physical system under a shift of zero-energy reference point. The effect of the approach is that the time-of-arrival operator plays the role of the generator of energy shift. Within the framework of field quantization, the total time-of-arrival is the sum of all the single event-of-arrival contributions, and contains zero-point quantum fluctuations because our quantum clock follow the laws of quantum mechanics.

In comparison, quantum field theory tells us that the total energy of a quantum field is the sum of all single-particle contributions, while Eq. (35) shows that the total time-of-arrival is the sum of all single-event contributions. With the zero-point fluctuations of total energy in quantum field theory, people have provided a satisfactory interpretation for spontaneous radiation phenomena and theoretically predicted the Casimir effect. In our theory, correspondingly, there are zero-point fluctuations of time from Eq. (35). The roles of these fluctuations in physical world remain to be explored.

**Acknowledgments**

The first author (Z. Y. Wang) would like to thank professors Erasmo Recami for his helpful discussions and J. G. Muga for his useful comments. Project supported by the National Natural Science Foundation of China (Grant No. 60671030) and by the Scientific

Research Foundation for the Introduced Talents, UESTC (Grant No. Y02002010501022).

**Appendix A**

For example, let us prove that

$$\phi_{xs}(p) = (p^2/\sqrt{p^2+m^2})^{1/4} u(p,s) \exp(-\mathrm{i}px)/(2\pi)^{1/2}, \text{ for } T = -T_x = -xE_p/p, \quad (5)$$

satisfy

$$\hat{T}_{\text{Dirac}}(\hat{x},\hat{p})\phi_{xs}(p) = -T_x\phi_{xs}(p), \quad (a1)$$

i.e.,

$$\hat{T}_{\text{Dirac}}(\hat{x},\hat{p})\phi_{xs}(p) = [\frac{1}{p}(\alpha_1 p + \beta m)(-\mathrm{i}\frac{\partial}{\partial p}) + \mathrm{i}\beta\frac{m}{2p^2}]\phi_{xs}(p) = -(xE_p/p)\phi_{xs}(p), \quad (a2)$$

let $\hat{T}_{\text{Dirac}}(\hat{x},\hat{p})\phi_{xs}(p) = L + (\mathrm{i}\beta m/2p^2)\phi_{xs}(p)$, where

$$L = [\frac{1}{p}(\alpha_1 p + \beta m)(-i\frac{\partial}{\partial p})]\phi_{xs}(p) = A + B + C, \tag{a3}$$

$$\begin{cases} A = [-i\frac{\partial}{\partial p}(p^2/p^2+m^2)^{1/4}]\frac{1}{p}(\alpha_1 p + \beta m)u(p,s)\exp(-ipx)/(2\pi)^{1/2} \\ B = (p^2/p^2+m^2)^{1/4}\frac{1}{p}(\alpha_1 p + \beta m)[-i\frac{\partial}{\partial p}u(p,s)]\exp(-ipx)/(2\pi)^{1/2} \\ C = (p^2/p^2+m^2)^{1/4}\frac{1}{p}(\alpha_1 p + \beta m)u(p,s)[-i\frac{\partial}{\partial p}\exp(-ipx)/(2\pi)^{1/2}] \end{cases} \tag{a4}$$

Applying Eqs. (7) and (8), we have ( $E_p^2 = p^2 + m^2$ )

$$\frac{\partial}{\partial p}(\frac{p^2}{p^2+m^2})^{1/4} = \frac{m^2}{2E_p^2}\frac{1}{p}(\frac{p^2}{p^2+m^2})^{1/4}, \quad \frac{\partial}{\partial p}u(p,s) = (\frac{m^2}{2E_p^2})\alpha_1\beta u(p,s), \tag{a5}$$

Substituting (a5), $\partial\exp(-ipx)/\partial p = -ix\exp(-ipx)$ and $(\alpha_1 p + \beta m)u(p,s) = E_p u(p,s)$

into (a4), we have

$$A = -i\frac{m^2}{2E_p}\frac{1}{p^2}\phi_{xs}(p), \quad B = i\frac{m}{2E_p}\frac{1}{p}\alpha_1\beta\phi_{xs}(p), \quad C = -(\frac{E_p x}{p})\phi_{xs}(p). \tag{a6}$$

Then

$$\hat{T}_{\text{Dirac}}(\hat{x},\hat{p})\phi_{xs}(p) = A + B + C + (i\beta m/2p^2)\phi_{xs}(p) = D - (\frac{E_p x}{p})\phi_{xs}(p), \tag{a7}$$

where

$$\begin{aligned} D &= A + B + (i\beta m/2p^2)\phi_{xs}(p) \\ &= i(\frac{m}{2p^2}\beta - \frac{m^2}{2E}\frac{1}{p^2} + \frac{m}{2E}\frac{1}{p}\alpha_1\beta)u(p,s)(\frac{p^2}{p^2+m^2})^{1/4}\exp(-ipx)/(2\pi)^{1/2} \\ &= iF(\frac{E+m}{2E})^{\frac{1}{2}}(\frac{p^2}{p^2+m^2})^{1/4}\exp(-ipx)/(2\pi)^{1/2} \end{aligned} \tag{a8}$$

where

$$F = (\frac{m}{2p^2}\beta - \frac{m^2}{2E}\frac{1}{p^2} + \frac{m}{2E}\frac{1}{p}\alpha_1\beta)\begin{pmatrix} \eta_s \\ \frac{\sigma_1 p}{E+m}\eta_s \end{pmatrix} \tag{a9}$$

Using $\sigma_1^2 = 1$, and

$$\alpha_1 = \begin{pmatrix} 0 & \sigma_1 \\ \sigma_1 & 0 \end{pmatrix}, \quad \beta = \begin{pmatrix} I & 0 \\ 0 & -I \end{pmatrix}, \quad I = \begin{pmatrix} 1 & 0 \\ 0 & 1 \end{pmatrix}, \tag{a10}$$

one has

$$F = \frac{m}{2p^2}\begin{pmatrix} \eta_s \\ -\dfrac{\sigma_1 p}{E+m}\eta_s \end{pmatrix} + \frac{m}{2E}\frac{1}{p}\begin{pmatrix} \dfrac{-p}{E+m}\eta_s \\ \sigma_1 \eta_s \end{pmatrix} - \frac{m^2}{2E}\frac{1}{p^2}\begin{pmatrix} \eta_s \\ \dfrac{\sigma_1 p}{E+m}\eta_s \end{pmatrix} = 0. \tag{a11}$$

Because $F = 0 \Rightarrow D = 0$, using Eq. (a7), one has

$$\hat{T}_{\text{Dirac}}(\hat{x},\hat{p})\phi_{xs}(p) = -T_x\phi_{xs}(p) = -(xE_p/p)\phi_{xs}(p). \tag{a12}$$

Then Eq. (5) satisfy Eq. (a1).